\documentclass[journal,english,twocolumn,10pt,letterpaper]{IEEEtran}

\usepackage{babel}
\usepackage[utf8]{inputenx}
\usepackage[T1]{fontenc}
\usepackage{graphicx}
\usepackage{rotating}
\usepackage{color}
\usepackage{url}
\usepackage[cmex10]{amsmath}
\usepackage{amssymb}
\usepackage{balance}
\usepackage{tikz}
\usepackage[colorlinks,bookmarksopen,bookmarksnumbered,citecolor=red,urlcolor=red]{hyperref}

\IEEEoverridecommandlockouts

\newcommand{\Esp}{\mathrm{E}}
\newcommand\copyrighttext{%
  \footnotesize \textcopyright 2015 IEEE. Personal use of this material is permitted.
  Permission from IEEE must be obtained for all other uses, in any current or future 
  media, including reprinting/republishing this material for advertising or promotional 
  purposes, creating new collective works, for resale or redistribution to servers or 
  lists, or reuse of any copyrighted component of this work in other works. 
  DOI: \href{http://dx.doi.org/10.1109/LWC.2015.2472398}{10.1109/LWC.2015.2472398}
}
\newcommand\copyrightnotice{%
\begin{tikzpicture}[remember picture,overlay]
\node[anchor=south,yshift=10pt] at (current page.south) {\fbox{\parbox{\dimexpr\textwidth-\fboxsep-\fboxrule\relax}{\copyrighttext}}};
\end{tikzpicture}%
}

\begin{document}

\title{A Self-Tuning Receiver-Initiated MAC Protocol for Wireless Sensor Networks}

\author{Miguel~Rodríguez-Pérez,~\IEEEmembership{Member,~IEEE}, Sergio~Herrería-Alonso,\\ Manuel
  Fernández-Veiga,~\IEEEmembership{Senior Member,~IEEE,} 
  and~Cándido~López-García%
  \thanks{The authors are with the Telematics Engineering Dept., Univ.~of
    Vigo, 36$\,$310~Vigo, Spain. email:~\protect\url{miguel@det.uvigo.es} (M. Rodríguez-Pérez).}}

\maketitle 
\copyrightnotice

\begin{abstract}
  Receiver-initiated medium access control protocols for wireless sensor
  networks are theoretically able to adapt to changing network conditions in a
  distributed manner. However, existing algorithms rely on fixed beacon rates
  at each receiver. We present a new received initiated MAC protocol that
  adapts the beacon rate at each receiver to its actual traffic load. Our
  proposal uses a computationally inexpensive formula for calculating the
  optimum beacon rate that minimizes network energy consumption and, so, it
  can be easily adopted by receivers. Simulation results show that our
  proposal reduces collisions and diminishes delivery time maintaining a low
  duty cycle.
\end{abstract}

\begin{IEEEkeywords}
  Sensor Networks, Received Initiated MAC, Rate adaptation
\end{IEEEkeywords}

\section{Introduction}
\label{sec:introduction}
Medium access control (MAC) protocols for wireless sensor networks (WSN) aim
to optimize power consumption in addition to other traditional design concerns
of MAC schemes, as throughput or delay. Actually, the objective is that
transmitters and receivers (listeners) agree on a rendezvous point that
minimizes the time during which their radios are active. MAC protocols for
WSNs can be broadly divided in two groups: globally synchronized, where a
centralized entity orchestrates the transmission opportunities for each
device, and asynchronous protocols where nodes act independently of any
central entity. Asynchronous schemes can support local adaptation to varying
traffic load in different parts of the network without relying on a global
transmission schedule. Within this class, there exist both sender- and
receiver-initiated MAC protocols. The latter family, known to be generally
more network efficient~\cite{lin04:_power,sun08:_ri_mac,fafoutis15:_receiv},
is the focus of this work.

In receiver-initiated MAC protocols, transmissions are regulated by receiving
nodes. Whenever a sender has data for a neighbor node, it monitors the channel
waiting for the reception of a short beacon broadcasted by the desired
receiver. The beacon signs that the receiver is currently active and listening
the channel, so the transmission can proceed. During this waiting time,
sending nodes have to keep their radios on to detect the beacon transmission,
wasting energy. Another waste of energy happens if two close senders receive a
beacon in the same time interval, as their transmission will collide, even if
the beacons came from different receivers. PW-MAC~\cite{tang11:_pw_macy} deals
with both problems by transmitting the beacons at times predictable by the
senders. The receivers randomize their inter-beacon period with a
pseudo-random number generator, whose parameters are sent in every beacon.
Thus, after hearing a first beacon, a sender can find out all the future
beacon transmission times and avoid to monitor the channel continuously. Thus,
the radio of the sender can be turned on just a short time before the next
expected beacon transmission.

Clearly, PW-MAC cannot adapt to a variable traffic load unless its randomized
inter-beacon times respond to changes in time and space. In multi-hop sensor
networks the load varies with time and also with location, since nodes closer
to the sinks aggregate traffic from farther senders. An energy-efficient MAC
protocol should modulate the beacon interval and preserve a duty cycle as low
as that in PW-MAC.


In this paper, we build on PW-MAC and present a receiver-initiated MAC
protocol for WSNs able to generate the beacons in a predictable yet adaptive,
load-dependent way. We also derive the optimum beacon rate that minimizes the
energy consumption in the network, given a traffic load. The optimum rate
follows a simple formula which can be easily implemented in computationally
constrained sensors. The rest of the paper is organized as follows.
Section~\ref{sec:rate-adaption-at} presents a complete description of our
enhancements to existing receiver-initiated MAC algorithms. We present
experimental results in Section~\ref{sec:results}. Finally, the conclusions
are summarized in Section~\ref{sec:conclusions}.

\section{Protocol Description}
\label{sec:rate-adaption-at}

As stated in the Introduction, our work builds on the
PW-MAC~\cite{tang11:_pw_macy} protocol for its basic operation. Like in other
receiver-initiated MAC protocols, the receivers transmit beacon frames so that
neighbor nodes with pending traffic can send their data. PW-MAC uses a
pseudo-random sequence known to every node in the network for scheduling its
beacon transmission times. De-synchronizing the beacons reduces collisions,
while making their actual transmission time predictable avoids increasing the
duty cycle. Notice that, for randomizing the beacons, a pseudo-random
generator of high statistical quality is not necessary, and a description of
its seed can occupy very few bytes in the frame header. For instance, PW-MAC
uses just six bytes in the beacon to transmit the parameters of the linear
congruential generator (LCG) used to generate the inter-beacon times.

If adaptation to different traffic incoming rates is desired, the beacon rate
cannot remain static. In particular, when the traffic load increases, the
number of beacons generated in a given time interval must be increased adding
new extra beacons. However, changing the beacon frequency proportionally to
the traffic load is naive and fails to work properly, because all the nodes in
the network ---active or currently inactive--- should be notified
synchronously. Otherwise, non-listening neighbors (neighbors with no data for
the receiver when the beacon is transmitted) would be unaware of the changes
and would fail to meet at future rendezvous points. Consequently, the extra
beacons (in the following, sub-beacons) must be independent of the regular or
\emph{primary} beacons.

One way to schedule the extra beacons is to divide the interval between two
primary beacons in $f$ subintervals, where $f$ is a speeding factor, and
transmit extra beacons between these subintervals. The speeding factor can be
announced in the regular beacons without disturbing the operations of
non-listening nodes. Since changes to $f$ are likely to be sporadic, the nodes
could assume that the speeding factor stays in effect indefinitely. However,
this assumption poses a problem when the speeding factor changes (even if it
is incremented): the boundaries between the subintervals change wildly, and
the nodes that lack updated information cannot thereafter predict correctly
the transmission times of the sub-beacons.

In the rest of this Section, we describe how to modify PW-MAC to accomplish
rate-adaptation. We will however ignore the details concerning the clock
adjustments among nodes and the collision resolution protocol, for their
behavior in PW-MAC applies unmodified to our proposal.

\subsection{Generating Predictable Sub-beacons}
\label{sec:gener-pred-extra-1}

We have resorted to fix the transmission times of the sub-beacons, but make
their actual transmissions dependent on the speeding factor. To this end, let
us divide the inter-beacon interval into fixed intervals in which a sub-beacon
\emph{could} potentially be transmitted. The actual transmission of each
individual sub-beacon is going to depend on the speeding factor $f$ and it
will be monotone, i.e., if a sub-beacon is transmitted for a given factor $f$,
then it will be also transmitted when the speeding factor increases. This
assumption allows uninformed senders to predict the future sub-beacons even in
case that the traffic load increases.

\begin{figure}
  \centering
  \resizebox{.5\textwidth}{!}{\input{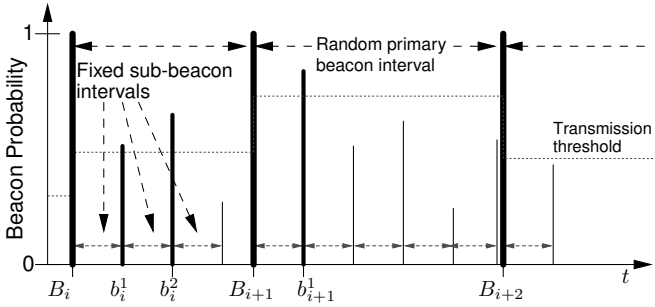}}
  \caption{Predictable sub-beacon generation procedure. Actually transmitted
    sub-beacons appear in bold lines. In the example $n_{\mathrm b}=4$. For $t
    \in [B_i, B_{i+1})$, $f=3$ so
    $th_{\mathrm b}=0.5$. For $t \in [B_{i+1}, B_{i+2})$, $f=2$ so
    $th_{\mathrm b}=0.75$.}
  \label{fig:beacon-generation}
\end{figure}
Specifically, each sub-beacon is associated with a sending probability
computed with an independent LCG with initial seed equal to the current value
of the random generator of the inter-beacon times. Additionally, a sub-beacon
transmission threshold $th_{\mathrm b}$ is computed so that only sub-beacons
whose associated sending probability, as per the LCG, is greater that
$th_{\mathrm b}$ are actually transmitted. This threshold is calculated so
that, on average, $f-1$ sub-beacons are actually transmitted, so
$th_{\mathrm b }=1-(f-1)/n_{\mathrm b}$, where $f$ stands the speeding factor and
$n_{\mathrm b}$ is the average number of sub-beacons between two regular
beacons. The actual transmission time $b_i^j$ of the $j$-th sub-beacon in the
$i$-th interval, for $j = 1, \dots, k_i$, is
$B_i + j \cdot \Delta_B /n_{\mathrm b}$ where $B_i$ denotes the transmission
time of the previous regular beacon, $\Delta_B = \Esp[B_{i + 1} - B_i]$ is the
average cycle length, and $k_i$ is such that
$B_i + k_i \cdot \Delta_B / n_{\mathrm b} < B_{i+1} \leq B_i + (k_i + 1)\Delta_B /
n_{\mathrm b}$.
The overall scheme is depicted in Fig.~\ref{fig:beacon-generation} that shows
the variation of the transmission threshold, the fixed length inter
sub-beacons length, and how only sub-beacons with an associated probability
greater than the threshold are actually transmitted, represented with bold
lines.

\subsection{Choosing the Beacon Rate}
\label{sec:choos-right-beac}

Receivers can use the generation of sub-beacons to add more transmission
opportunities for the senders. There remains, however, the problem of
determining how many sub-beacons should be added between each pair of regular
beacons.

We propose to maintain an estimation of the incoming rate $\lambda$ at each
receiver in order to choose the optimum number of extra beacons. Such estimate
could be calculated, for instance, with an exponential moving average of the
number of successful receptions in each regular cycle, but other methods are
possible. Assuming that the number of packets per cycle is known, we can
proceed to calculate the beacon multiplication factor $f$. Note that
$f = \lambda$ is generally not the best value, as there is an asymmetry
between the cost of a failed transmission (collision) and that of an
unanswered beacon.

\subsubsection{Deriving the Optimum Beacon Rate}
\label{sec:sett-corr-beac}

For obtaining the optimum multiplication factor with respect to energy
consumption, we will first calculate the average amount of energy wasted by a
given receiver and all its neighbor nodes with backlogged traffic to the
receiver in a cycle, where a cycle is the time between two consecutive primary
beacons. There are two sources for such energy expenditure: transmission
collisions and unused beacons. Let us consider collisions first.

When there occur two or more simultaneous transmissions to a single receiver a
collision happens and the energy $E_{\mathrm{b}}$ employed to transmit the
beacon is wasted.\footnote{For the analysis, we assume a simplified worst case
  model where receivers cannot recover any frame whenever a collision
  happens.} Furthermore, the energy used by each sender listening while
waiting for the beacon $E_{\mathrm{w}}$ and to actually transmit the data
frame $E_{\mathrm{tx}}$ is also squandered. So, the amount of energy wasted by
collisions in a cycle is
\begin{equation}
  \label{eq:ecol-1}
  E^1_{\mathrm{col}}(\lambda) = E_{\mathrm{b}} + (E_{\mathrm{w}}+E_{\mathrm{tx}})
  \sum_{i=2}^\infty F(\lambda, i),
\end{equation}
where $\lambda$ is the average incoming rate at the receiver, measured in
packets per cycle, $F(\lambda, i)$ is the probability mass function of the
arrival distribution for the receiving node, and $i$ is the number of
simultaneous transmissions in the cycle.

When the sub-beacons are in use, the cycle is split in $f$ sub-cycles and
the total lost energy in the whole cycle is
\begin{equation}
  \label{eq:ecol-f}
  E^f_{\mathrm{col}}(\lambda) = f E^1_{\mathrm{col}}(\lambda/f).
\end{equation}

However, as we reduce the cycle length, we are also increasing the beacon
rate, and some energy is consumed sending unnecessary beacons. The amount of
wasted energy due to that reason is given by
\begin{equation}
  \label{eq:e-unused-beacon}
  E^f_{\emptyset}(\lambda) = f E_{\mathrm{b}} F(\lambda/f, 0).
\end{equation}
So, the optimum multiplication factor $f$ is obtained when
$E^f_B(\lambda) = E^f_{\mathrm{col}}(\lambda) + E^f_{\emptyset}(\lambda)$ is
minimum.

\subsubsection{Optimum Beacon Rate for Poissonian Traffic}
\label{sec:optimum-beacon-rate}

We can particularize the previous result for the special case of a Poisson
distribution. Although in the edges of the network the distribution of time
between frames in a given node is arbitrary, as we move towards the network
core and traffic is aggregated the frame arrivals converge to a Poisson
process, by virtue of the Palm-Khintchine theorem. Moreover, as we are mostly
interested in calculating the number of extra beacons needed to adapt to this
aggregated traffic, we can reasonably assume that nodes with extra beacons
will receive, in fact, Poissonian traffic.\footnote{Albeit this assumption is
  invalid for high loads as the re-transmitted traffic is correlated, our
  method is able to accommodate higher traffic rates before collisions happen
  and thus the assumption remains valid for usual workloads.} So, substituting
$F(\lambda, k) = \frac{\lambda^k}{k!} \mathrm{e}^{-\lambda}$
in~\eqref{eq:ecol-f} and~\eqref{eq:e-unused-beacon} we get
$E^f_B(\lambda) = f E_{\mathrm{b}} + \lambda (E_{\mathrm{w}} +
E_{\mathrm{tx}}) +\mathrm{e}^{-\lambda/f} (f E_{\mathrm{b}} -
\lambda(E_{\mathrm{w}} + E_{\mathrm{tx}})).$
Now, finding the optimum $f$ is just a matter of deriving $E_B(\lambda, f)$ with respect to
$f$ and solving
\begin{equation}
  \label{eq:exact-f}
  \frac{\partial}{\partial f}E_B^f(\lambda)=E_{\mathrm{b}} (f + \mathrm{e}^{\lambda/f} f + \lambda)f -
  \lambda^2(E_{\mathrm{w}} + E_{\mathrm{tx}}) = 0.
\end{equation}

\begin{figure}
  \centering
  \includegraphics[width=.5\textwidth]{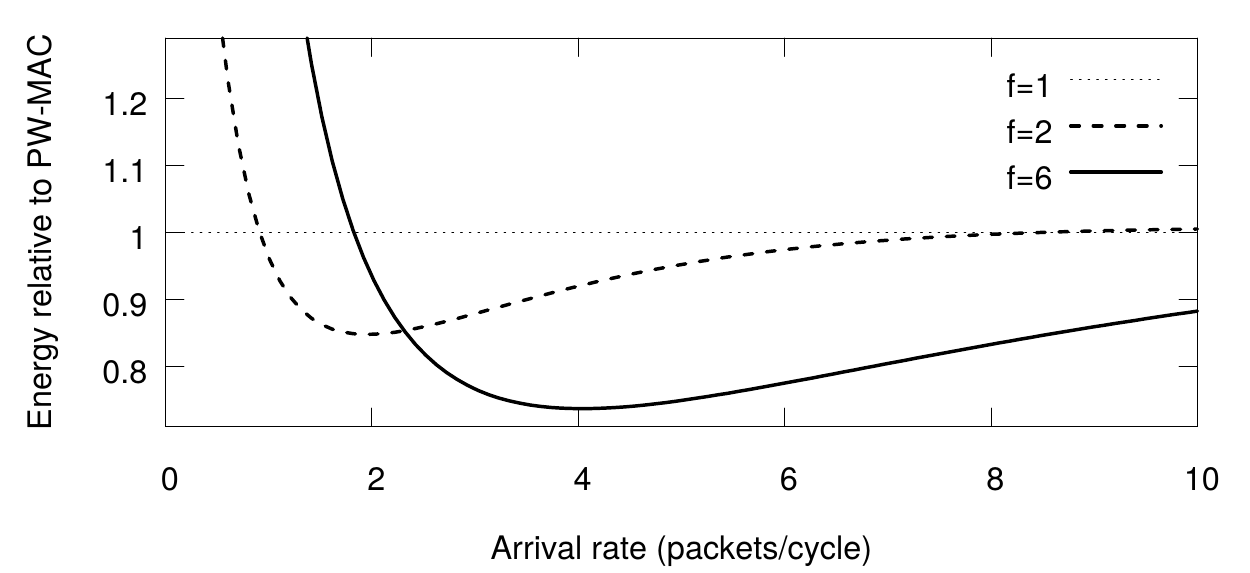}
  \caption{Energy waste, $E_B^f(\lambda)$, for different acceleration factors
    $f$ and varying arrival rate $\lambda$ compared to PW-MAC ($f=1$).}
  \label{fig:profit}
\end{figure}

Unfortunately, there is no closed form for $f$ and resorting to numerical
methods is out of the question for resource-constrained sensors.
For solving~\eqref{eq:exact-f}, we can take advantage of the fact that the
minimum energy is consumed for values of $f$ close to $\lambda$, measured in
packets per cycle. This is illustrated in~Fig.~\ref{fig:profit}, which shows
the total energy waste $E_B^f(\lambda)$ versus $f$ for a range of traffic
loads. The plotted values are normalized against PW-MAC, that corresponds with
$f=1$, that is, without using sub-beacons.
Therefore, we approximate $E_B^f(\lambda)$ by its second order Taylor series
expansion around $\lambda$, assuming that the optimum $f$ value will be near
the incoming rate
\begin{align}
  \label{eq:energy-beacon-taylor-2}
  E_B^f(\lambda) &\approx \lambda\frac{(1+\mathrm{e}) E_{\mathrm{b}} - (1-\mathrm{e}) (E_{\mathrm{w}} +
                    E_{\mathrm{tx}})}{\mathrm{e}}\nonumber\\
                  &+\frac{(E_{\mathrm{w}} + E_{\mathrm{tx}}+E_{\mathrm{b}})(f-\lambda)^2}{2\mathrm{e}\, \lambda}\nonumber\\
                  &-\frac{\left(\frac{E_{\mathrm{w}} +
                    E_{\mathrm{tx}}
                    -2 E_{\mathrm{b}}}{\mathrm{e}}+E_{\mathrm{b}}\right) (f-\lambda)}
                    {2\mathrm{e}\, \lambda},
\end{align}
derive~\eqref{eq:energy-beacon-taylor-2} with respect to $f$ and solve. The
final result is
\begin{equation}
  \label{eq:taylor-f-1}
f^\ast = \lambda\frac{2 (E_{\mathrm{w}} + E_{\mathrm{tx}}) - (1+\mathrm{e})E_{\mathrm{b}}}{E_{\mathrm{b}} + E_{\mathrm{w}} + E_{\mathrm{tx}}}
\end{equation}
where all the terms are known in advance except $\lambda$, which is directly
measured by the nodes.

\section{Results}
\label{sec:results}

We have experimentally verified that EH-MAC improves the
original PW-MAC in the key performance metrics: delivery delay, sensor duty
cycle, delivery reliability and collision probability. To this end, we have
simulated a representative set of deployment scenarios.\footnote{The code for
  the simulations is available for download at
  \url{https://migrax.github.io/EH-MAC}.} To serve as a reference point, we
also tested the performance of the original RI-MAC~\cite{sun08:_ri_mac}
algorithm.

The parameters used in the numerical experiments were as follows. The main
beacon interval at each node independently follows an uniform distribution
between 500$\,$ms and 1500$\,$ms as in the original PW-MAC. The maximum
sub-beacon rate has been capped to one sub-beacon every 100$\,$ms, so the
maximum long-term sustainable rate at the receiver is 10$\,$packets$/$s. Each
EH-MAC receiver selects the beacon rate according to~\eqref{eq:taylor-f-1} and
estimates $\lambda$ with a moving window of the last 15 inter-arrival
times. Senders have been set to wake up 10$\,$ms in advance to the predicted
beacon transmission to account for the clock drift. Finally, the packets are
128$\,$bytes long and the transmission rate is 250$\,$kb$/$s, whereas beacons
are 60$\,$bits long.
All the simulations were carried out on a 100$\,$m$\times$100$\,$m field where
nodes were randomly placed according to a spatial Poisson distribution of a
given homogeneous density. To account for the fact that most traffic in the
network goes towards a single sink node, a random node was chosen as the
delivery destination. Traffic was then transmitted to it via multiple-hops
using a greedy routing algorithm.
The transmission distance was set to 35$\,$m. Every node generates traffic as
a Poisson process with the same fixed rate. Every experiment was run
for 1000 seconds and repeated 100 times varying the seed of the random
generator of the simulator and the chosen sink. 95$\,$\%~confidence intervals
of every metric were also calculated.

\begin{figure}
  \centering
  \includegraphics[width=.5\textwidth]{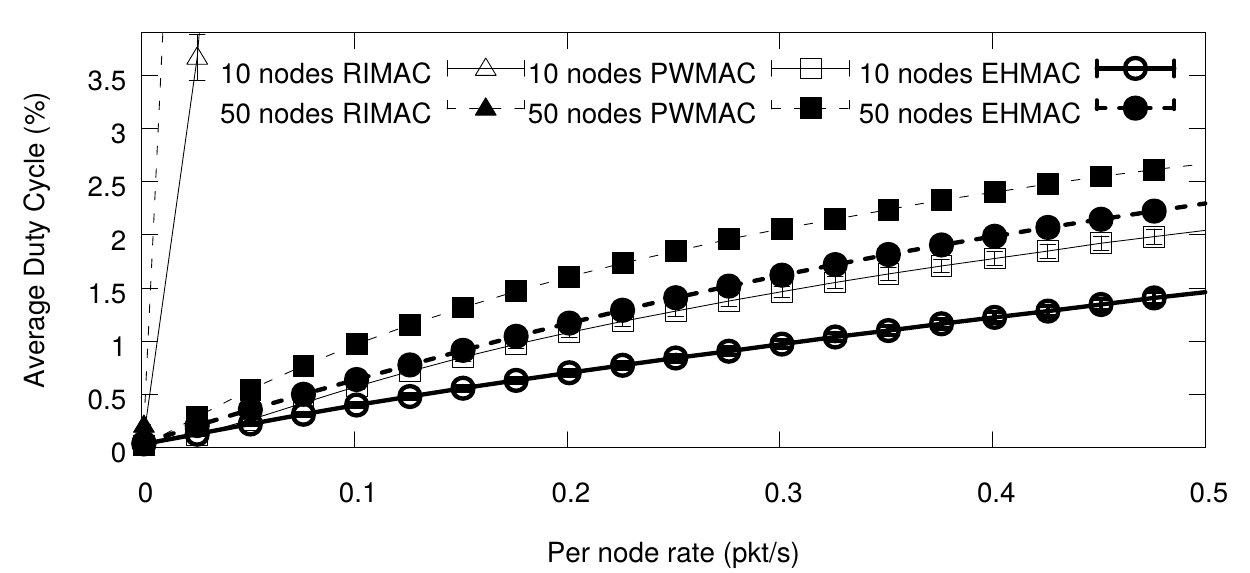}
  \caption{Duty cycle as a function of load.}
  \label{fig:duty-cycle}
\end{figure}
Figure~\ref{fig:duty-cycle} shows average duty cycle of the network nodes for
the three MAC algorithms compared and two representative node densities. The
average duty cycle, being the percentage of time that sensors have to keep
their radios on, is ultimately responsible for the MAC protocol power
consumption. The traffic rate was varied between 0 and 0.5$\,$packets$/$s.
Please note that this corresponds with a 5$\,$packets$/$s rate at the receiver
for the 10 nodes scenario and 25$\,$packets$/$s for that with 50 nodes, so
collisions are likely to occur near the sink. We clearly see that both PW-MAC
and EH-MAC significantly outperform the original RI-MAC, while, at the same
time, EH-MAC has an slightly better performance than PW-MAC.

\begin{figure}
  \centering
  \includegraphics[width=.5\textwidth]{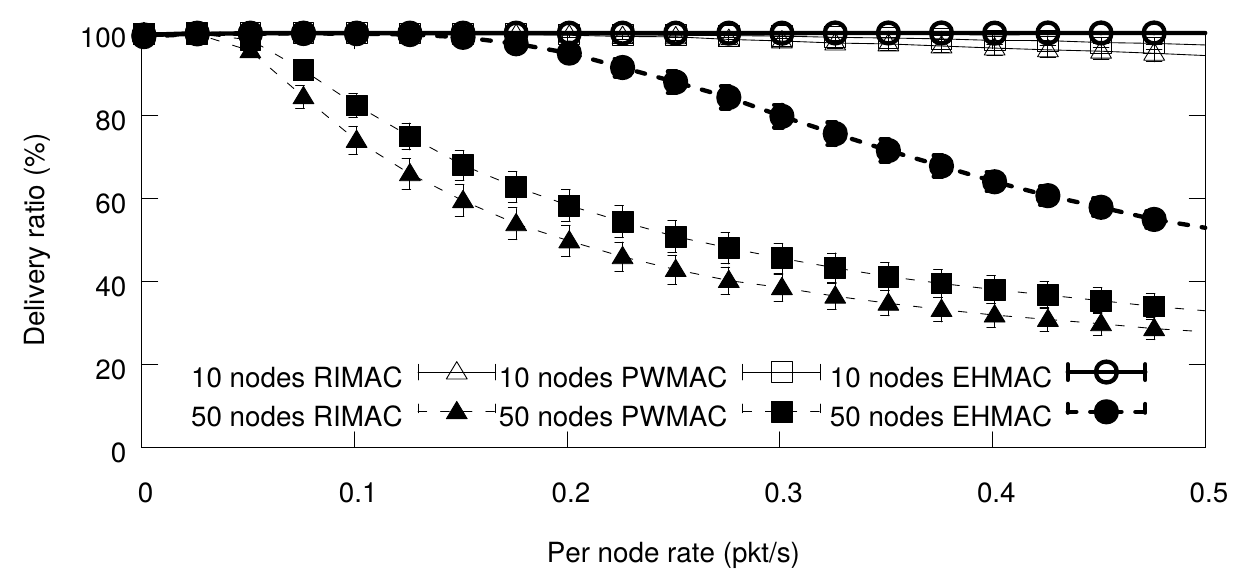}
  \caption{Delivery reliability as a function of load.}
  \label{fig:delivery-ratio}
\end{figure}
The successful delivery ratio is plotted in Fig.~\ref{fig:delivery-ratio}. In
the ten nodes setting, we see that all the protocols achieve an almost perfect
reliability, although both RI-MAC and PW-MAC loose some packets at the highest
loads. EH-MAC, on the contrary, is able to increase the beacon rate to
accommodate the needs of the senders, achieving the perfect reliability even
at the highest loads. For the 50 nodes case, the results get worse. When the
rate reaches 0.02$\,$packets$/$s, the effective rate at the receiver reaches
1$\,$packet$/$s, the maximum both RI-MAC and PW-MAC were configured to
support, and packets start to be discarded. In contrast, EH-MAC is stable
until the generation rate reaches 0.2$\,$packets$/$s, one order of magnitude
greater. At that point, its performance also starts to diminish but, in any
case, it stays considerably better than that of PW-MAC.

\begin{figure}
  \centering
  \includegraphics[width=.5\textwidth]{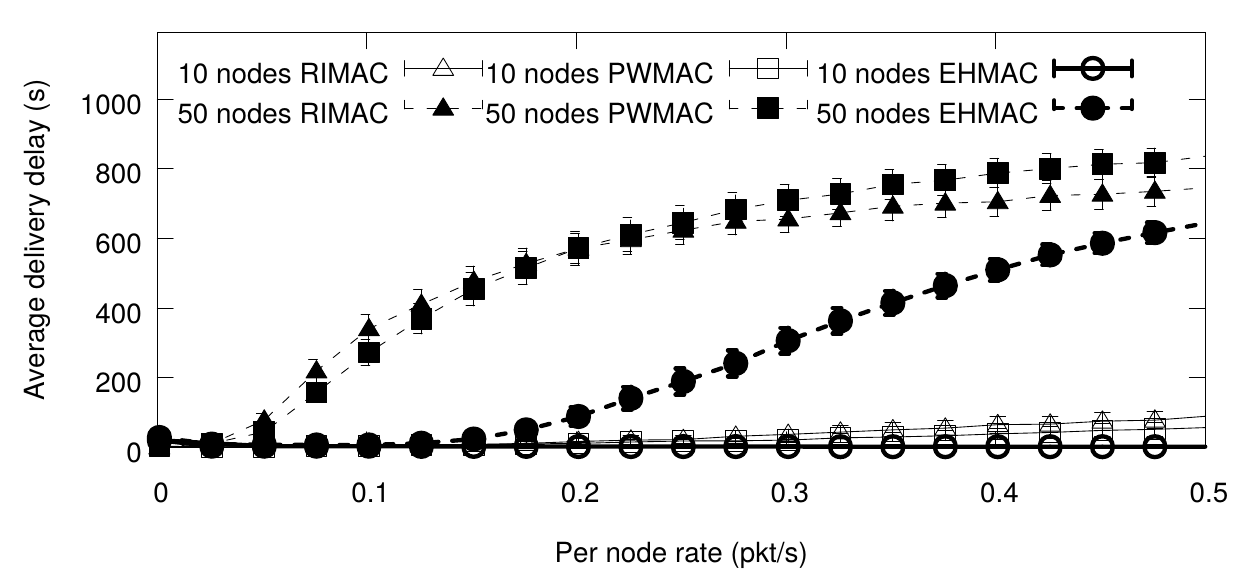}
  \caption{Average packet delivery delay as a function of load.}
  \label{fig:delay}
\end{figure}
The average packet delivery delay, represented in Fig.~\ref{fig:delay}, shows
a similar trend to the delivery ratio. For the 10 nodes case the three
algorithms show a good behavior, with slightly better results for our
proposal. The improvement is much greater in the 50 nodes scenario, where it
is clearly shown that EH-MAC copes with higher incoming rates until the
maximum is reached and the system cannot admit all the offered load. Note that
the maximum delay is about 900 seconds because the simulations were limited to
1000 seconds, so no delivered packet can show higher delays. Unfortunately,
this also gives the false impression that the delay asymptotically converges
to a maximum value, whereas in reality either the delay would keep increasing
unboundedly, or more packets would be lost.

\begin{figure}
  \centering
  \includegraphics[width=.5\textwidth]{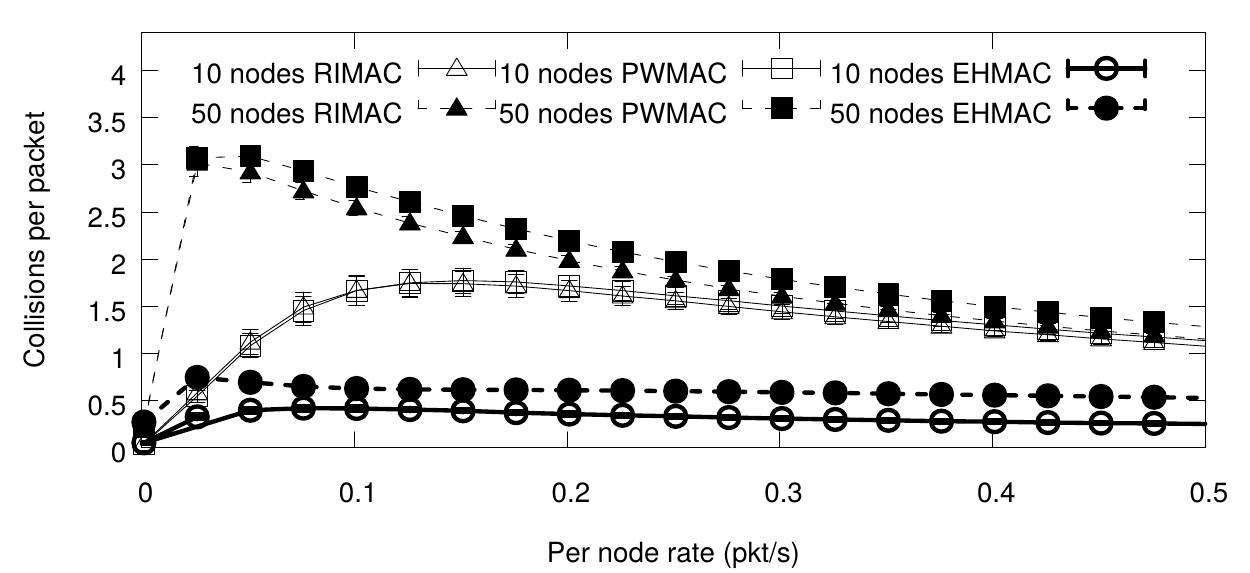}
  \caption{Number of collisions as a function of load.}
  \label{fig:collisions}
\end{figure}
The last figure shows the total number of collisions. Both RI-MAC and PW-MAC
behave similarly, with more collisions than our enhanced EH-MAC in both
settings, as shown in Fig.~\ref{fig:collisions}. But, as the load grows, the
number of collisions decreases. This is because for
high load many nodes are retransmitting collided packets,
ultimately enlarging the time receivers wait for packets. For the enhanced
algorithm, the effect is far less noticeable, because the algorithm adjusts
the beacon rate so as to fit the demand of the senders. Otherwise, for the
highest loads, collision retransmission comes up like in RI-MAC and PW-MAC.

\section{Conclusions}
\label{sec:conclusions}

We have proposed EH-MAC, a rate-adaptive mechanism for beacon generation in
received initiated MAC protocols. EH-MAC produces an optimum number of new
predictable transmission beacons while maximizing energy savings. The
generation of new beacons does not have a performance impact on those senders
unaware of them. We have compared via simulation the performance of EH-MAC
against PW-MAC, obtaining better results in all important
metrics.

\section*{Acknowledgments}
\label{sec:acknowledgment}
\addcontentsline{toc}{section}{Acknowledgments}

Work supported by the European Regional Development Fund (ERDF) and the
Galician Regional Government under agreement for funding the Atlantic Research
Center for Information and Communication Technologies
(\href{http://atlanttic.uvigo.es/en/}{AtlantTIC}).

\balance 

\bibliographystyle{IEEEtran}
\bibliography{IEEEabrv,biblio}

\end{document}